\documentclass[useAMS,referee]{biom}
\usepackage{graphicx}
\graphicspath{{pics/}}
\usepackage{setspace}
\usepackage{titlesec}

\usepackage{booktabs}
\usepackage{amssymb}
\usepackage{amsmath}
\usepackage{xcolor}

\usepackage{algorithm}
\usepackage{subfigure}

\usepackage{algpseudocode}
\usepackage[citestyle=authoryear,bibstyle=authoryear,doi,url,dashed=true]{biblatex}

\renewbibmacro*{in:}{%
  \iffieldundef{journaltitle}
    {}
    {\printtext{\intitlepunct}}}

\makeatletter
\newcommand{\@journal}{Biometrics}
\makeatother
  
\def\bSig\mathbf{\Sigma}

\usepackage{tikz}

\newcommand*{\tikzbulletred}{%
  \setbox0=\hbox{\strut}%
  \begin{tikzpicture}
  \fill[red] (0,0) circle (0.1cm);
\end{tikzpicture}%
}

\newcommand*{\tikzbulletgreen}{%
  \setbox0=\hbox{\strut}%
  \begin{tikzpicture}
  \fill[green] (0,0) circle (0.1cm);
\end{tikzpicture}%
}

\newcommand*{\tikzbulletyellow}{%
  \setbox0=\hbox{\strut}%
  \begin{tikzpicture}
  \fill[yellow] (0,0) circle (0.1cm);
\end{tikzpicture}%
}

\newcommand*{\tikzbulletblue}{%
  \setbox0=\hbox{\strut}%
  \begin{tikzpicture}
  \fill[blue] (0,0) circle (0.1cm);
\end{tikzpicture}%
}

\titlespacing*{\section}{0pt}{3pt}{0pt}

\titlespacing*{\subsection}{0pt}{1.5pt}{0pt}

\title[SBART Spatial Survival]{Analysis of spatially clustered survival data with unobserved covariates using SBART}

\author{Durbadal Ghosh$^{1,*}$, Debajyoti Sinha$^{1}$, Antonio Linero$^{2}$,  George Rust$^{1}$ \\
	{\small $^{1}$\it Florida State University;
	$^2$\it University of Texas at Austin
	}\\
	{\small \it *email: \texttt{dg20ea@fsu.edu}}}

\begin{document}

\date{{\it Received November} 2023. {\it Revised April} 2023.  {\it
Accepted March} someday.}

\volume{64}
\pubyear{2023}
\artmonth{December}

\label{firstpage}


%


\begin{abstract}
  Usual parametric and semi-parametric regression methods are inappropriate and inadequate for large clustered survival studies when the appropriate functional forms of the covariates and their interactions in hazard functions are unknown, and random cluster effects and cluster-level covariates are spatially correlated. We present a general nonparametric method for such studies under the Bayesian ensemble learning paradigm called Soft Bayesian Additive Regression Trees. Our methodological and computational challenges include large number of clusters, variable cluster sizes, and proper statistical augmentation of the unobservable cluster-level covariate using a data registry different from the main survival study. We use an innovative 3-step approach based on latent variables to address our computational challenges. We illustrate our method and its advantages over existing methods by assessing the impacts of intervention in some county-level and patient-level covariates to mitigate existing racial disparity in breast cancer survival in 67 Florida counties (clusters) using two different data resources. Florida Cancer Registry (FCR) is used to obtain clustered survival data with patient-level covariates, and the Behavioral Risk Factor Surveillance Survey (BRFSS) is used to obtain further data information on an unobservable county-level covariate of Screening Mammography Utilization (SMU).
\end{abstract}

\maketitle

\section{Introduction}
\label{s:intro}

Clustered survival responses frequently arise
in family studies, complex surveys, and multi-center clinical trials
because survival times are clustered within families, geographical regions/units, and clinical centers \parencite{H1}. 
Unlike existing estimation equations based marginal analysis of clustered survival data 
(e.g., Lin and Ying, 1997; Lin and Wei, 1992),
particularly for spatially clustered survival data, 
models with unobserved cluster-specific random effects 
allow more careful modeling of the within-cluster association.
The most popular methods for such random cluster effects are the so-called shared frailty models \parencite{Oakes82, Aalen94} under the assumption of proportional hazards \parencite{Cox72} for the covariate effect as well as the random cluster effect, called frailty.  
See Hougaard (2000) and Ibrahim et al. (2001) for reviews of frequentist and Bayesian approaches for clustered survival data using the shared frailty models, which usually assume independence among random cluster effects (frailties). However, 
our motivating survival study uses the Florida Cancer Registry (FCR), a state-wide registry of $n=76,174$ breast cancer patients accrued during 2004-2016 from 67 FL counties (clusters) with the availability of only the name of the county of each patient (hence, each county being an areal unit). 
So, these random cluster effects on survival (frailties) are expected to be spatially associated.
Like many other studies in the era of big data, 
the cluster sizes in FCR
are highly variable, with a median of 503, a minimum of 1 (Liberty County), and a maximum of 6,234 patients (Broward County). 
The results of a separate analysis of survival data from each county would produce very unreliable conclusions for small countries with a moderate number of patients. 
Recently, there has been a massive increase in such studies with spatial locations of subjects \parencite{Li02, Taylor2015, Banerjee16, Zhou17}.  
Unlike most of these articles dealing with point-reference spatial survival times, our method deals with spatial association
among unknown cluster/county effects by assuming a Conditional Auto Regressive (CAR) structure \parencite{Besag74} among areal (lattice) units.

One major challenge of our study is that 
one important county-level covariate, Screening-Mammography 
Utilization (in short, SMU), is not available from the FCR database even though it provides multiple other subject-level covariates. 
Our challenge of statistically principled augmentation of true unobservable SMU proportions within our Bayesian analysis is somewhat related to the covariate measurement error problem (Zucker, 2005; Sinha and Ma, 2014), except that
we utilize within our analysis another study, the Behavioral Risk Factor Surveillance System (BRFSS), a nationwide telephone survey including the utilization of cancer prevention services. 
So, the screening mammography survey data from BRFSS also depend on these unobservable SMU proportions. 
Instead of using any plug-in estimators of SMU proportions based on BRFSS data alone, we incorporate both BRFSS and FCR studies within the same joint posterior and use the CAR model \parencite{Besag74} to account for the spatial association among true unobservable SMU proportions of 67 counties.

Most existing methods for spatially correlated survival data use fully parametric or semiparametric survival models such as accelerated failure time (AFT) and proportional hazards \parencite{Cox72} models. 
In practice, the true functional forms of these functions of interest are possibly more complex than the assumed semiparametric regression model, mainly because the true functional forms and nature of interactions among the covariates are usually unknown.
From standard residual analysis, it is difficult to judge whether the semiparametric modeling assumptions are either valid or adequate to explain the covariate effects of such complex studies.
Especially for studies with large numbers of clusters and at least some large clusters, observed data may conflict with the restrictive modeling assumptions of popular parametric and semiparametric models.

Some of the recent nonparametric approaches to accommodate such complex survival regression methods include Boosting proportional hazards (Li and Luan, 2005), Bagging \parencite{Hothorn04}, Random Survival Forests (Ishwaran et al., 2008), 
and  Gaussian process for nonparametric covariate effects
\parencite{Fernandez16}.
Various tree-based nonparametric methods (Li, Linero, and Murray, 2022, and the references therein) for continuous and discrete responses (not survival response)  often work better than traditional nonparametric tools in
the presence of nonlinear regression effects and unknown order of interactions. 
Among these methods, the Bayesian Additive Regression Trees (BART)  particularly show promising prediction capability. An extension of BART, called Soft-BART (SBART),  can even adapt to the smoothness of the unknown regression function and avoid overfitting \parencite{Linero17}. For clustered survival data, 
a recently developed flexible method
using SBART \parencite{Basak21}  centers the prior of the nonparametric hazard on a parametric ``prior guess",  
accommodates time-dependent covariates and time-varying regression effects, and performs better than Random Survival
Forests and usual semiparametric Bayesian methods.
However, it does
not accommodate any unobserved cluster-level covariates and spatial associations among random cluster effects. 
Our other challenge is to obtain posterior samples
of the relevant parameters, predictions, and relevant model summaries to achieve our analysis goals of a study of disparity in breast cancer survival.

In Section~2, we present brief reviews of BART and SBART and then describe our model consisting of three sub-models: a hazard-based model for continuous and clustered survival responses, a spatial model for cluster/county effects, and a model for the BRFSS study data that also depends on the spatially correlated true SMU proportions of different clusters/counties. 
Section~3 describes our priors, including the SBART prior for our nonparametric hazard.  To deal with several computational challenges, including an intractable likelihood, two submodels sharing a common set of correlated parameters, a high number of clusters, and some very large cluster sizes, in Section~3, we present an efficient and novel 3 steps data-augmentation-based algorithm with the associated code in R to obtain  Markov Chain Monte Carlo (MCMC) samples from our joint posterior with easy to compute importance sampling weights. 
The simulation studies in Section~4 compare our method's performance with two existing frequentist methods. 
In Section~5, we compare our Bayesian analysis with the analysis based on competing methods of the Florida breast cancer study using different residual-based diagnostics and illustrate the practical advantages of our method through important analysis results. These include an ordering of the covariates based on their roles on the hazard, estimation of survival curves for patients with known covariates in different counties, and evaluation of expected lifetimes saved within 5 years and 10 years for a future patient if the racial disparities in certain covariates are mitigated. The last is of particular interest to policy makers. 
In Section~5, we conclude with some discussion about future directions.

\section{Models, likelihoods, and priors}

 \subsection{Brief review of BART and SBART:}
\begin{figure}
     \centering
     \includegraphics[width=.8\textwidth]{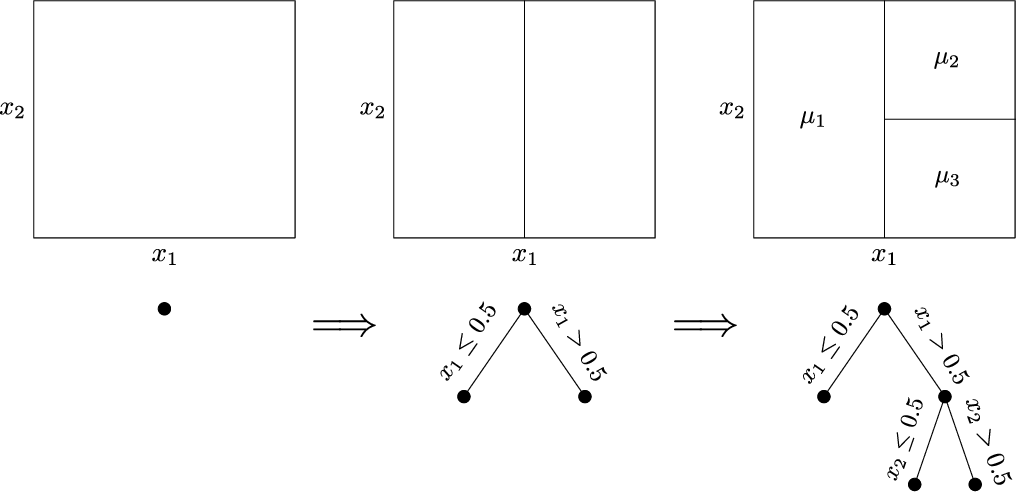}
    \caption{Schematic diagram illustrating how to draw from the prior on $(\tau, \mathcal M)$. For tree $\tau$, we first determine with probability $\gamma$ that the root node will be a branch;  then sample the splitting coordinate $j = 1$ and the cutpoint $C = 0.5$ for $x_j$. This process then iterates; the left child node is set to be a leaf node with probability $1 - \gamma/2^{2}$, and the right child is made another branch with probability $\gamma/2^2$. Eventually, this process terminates, and we independently sample a mean parameter $\mu$ for each of $J=3$ leaf nodes of the generated decision tree $\tau$.}
     \label{fig:TreeGrow}
\end{figure}

Even though Bayesian Regression Trees (BART) was originally developed for continuous response, BART now has been extended to binary response $Y$ with the nonparametric probit model
$Y \sim \operatorname{Bernoulli}[\Phi\{b_0(\mathbf{x})\}]$ (\cite{Chipman10}),
where $\Phi(\cdot)$ is the cdf of the standard normal density, 
and vector $\mathbf{x}$ is a $p$-dimensional covariates (including even categorical covariates). 
The unknown regression function $b_0:\mathbb{R}^p \to \mathbb{R}$ is a sum of $K$ regression trees $b_0(\mathbf{x}) = \sum_{k=1}^{K}g(\mathbf{x}; \tau_k, \mathcal{M}_k)$, 
where $\tau_k$ denotes the topology (splitting rules) of the decision tree $k$ and $\mathcal{M}_k = \{\mu_{k1}, \cdots, \mu_{k J_k}\}$ is the set of predictions for the $J_k$ terminal nodes (or resulting partitions of the covariate space) of the decision tree $\tau_k$. Here, each tree $g(\mathbf{x}; \tau_k, {\mathcal M_k}) =
     \sum_{\ell=1}^{J_k} \varphi_\ell(\mathbf{x}; \tau_k)  \mu_{k\ell}$ is a step function,
 where $\varphi_\ell(\mathbf{x}; \tau_k)$ is the indicator of whether $\mathbf{x}$ is associated to leaf $\ell$ of $\tau_k$. 
 For every branch node $b$ of $\tau_k$, a splitting rule of the form $[x_j \leq C_b]$ is assigned with $\mathbf{x}$ going down the left of the branch if the condition is satisfied and right of the branch otherwise.
 Following Chipman et al. (2010), unknown $\tau_k$ is assigned a branching process prior with sequentially employing regularized priors for terminating at a particular depth, for selecting a splitting variable at a nonterminal node, and for selecting the corresponding cut point of a splitting variable.
A schematic diagram in Figure~\ref{fig:TreeGrow} shows how the branching process prior generates a sample of the decision tree $\tau_k$ and, consequently, a partition of the covariate space and corresponding $\varphi_\ell(\mathbf{x}; \tau_k)$. 

 A schematic diagram in Figure~\ref{fig:TreeGrow} shows how the branching process prior generates a sample of the decision tree $\tau_k$ and, consequently, a partition of the covariate space and corresponding $\varphi_\ell(\mathbf{x}; \tau_k)$. 
 
\noindent The default common Gaussian prior $\mu_{kl} \stackrel{iid}{\sim} N(0,\sigma^2_\mu)$ for $\mu_{kl}$ with $\sigma_{\mu}=3/(2\sqrt{K})$ to
ensures that $\Phi\{b_0(\mathbf{x})\}$ is concentrated in the interval $(\Phi(-3),\Phi(+3))$. 
However, when the underlying response surface $b_0(\mathbf{x})$ is believed to be smooth, as is often the case in practice, prediction accuracy can be further improved theoretically, practically, and without compromising bias by extending the BART to SBART (Linero and Yang,
2018), which uses ``soft decision trees''   
via replacing the indicator function $\varphi_\ell(\mathbf{x}; \tau_k)$ with
$ \prod_{b \in \mathcal A_k(\ell)} 
     \psi(x_{j_b} ; C_b, \alpha_b)^{1-R_b} 
     \{1-\psi(x_{j_b}; C_b, \alpha_b)\}^{R_b}$, 
 where $A_k(\ell)$ denotes the collection of \emph{ancestor nodes} of leaf $\ell$ and $R_b$ is the indicator that the path from root to $\ell$ at branch $b$ goes right, and $\psi(x; c, \alpha)=[1 + \exp\{-(x - c) / \alpha\}]^{-1}$ is a continuous distribution function with location $c$ and scale $\alpha>0$. 
 The bandwidth parameter $\alpha \to 0$ corresponds to the usual BART model. 

Below, we describe how these are extended to address our survival submodel, followed by the spatial submodels for cluster effects and unknown cluster-level covariates.\\
\subsection{Nonparametric hazards model:}
Even though in our data analysis, we perform 
separate analyses for 6 strata indexed by combinations
of 2 races and 3 stages of cancer, we now suppress the obvious connection of our model parameters to the underlying stratum in our notation. 
For any particular stratum, the survival submodel for subject $j$ of the cluster (county) $i=1,\cdots,N$ has the hazard function
  \begin{equation}
  \lambda_{ij}(t\mid W_i, M_i;\ \mathbf{x}_{ij}) = \lambda_0(t)  \ W_i\  \Phi(b(t, M_i, {\mathbf x}_{ij}))
  \ \label{equ:model}
  \end{equation} 
given the cluster-specific random effect (frailty) $W_i$ and the true unobserved cluster-level covariate $M_i$. 
Here, \(\mathbf{x}_{ij} =(x_{ij1},...,x_{ijp}) \in \mathbb{R}^p\) is the vector of all observed (available in survival database FCR) patient-level covariates (examples include age and indicator of treatment delay). To ease the posterior computation, we scale all continuous covariates to have support $(0,1)$.  
The nonparametric hazard of (\ref{equ:model}) is centered on parametric baseline hazard $\lambda_0(\cdot)$. 
For example, if $b(t,M_i,{\mathbf x}_{ij})$ does not depend on $t$, then $\lambda_{ij}(t \mid W_i, M_i; {\mathbf x}_{ij}) = \lambda_0(t) \  W_i \Phi(b(M_i, {\mathbf x}_{ij}))$ is proportional over time to the baseline $\lambda_0(t)$. 
Without loss of generality, we assume constant \(\lambda_0(t)=\lambda_0\) because 
a fully nonparametric time-varying $b(t,\mathbf{x}_{ij},{M}_i)$ in (\ref{equ:model}) already involves $t$.
The unobserved cluster effect (frailty) $W_i$ accommodates the association within the cluster.
The hazard in (\ref{equ:model}) extends the hazard model of \cite{Basak21}) with SBART for unknown $b(t, M, \mathbf{x})$ to now accommodate spatial associations (explained later) between $W_i$ and among the unobservable $M_i$.
Under possible non-informative right-censoring,
 the likelihood contribution of the observed survival response \((y_{ij},\delta_{ij})\)  under the hazard function in (\ref{equ:model}) is 
 \begin{equation}
L_{ij}(W_i, \lambda_0,b; M_i)\propto \{\lambda_0 W_i\Phi(b(y_{ij}, M_i; \mathbf{x}_{ij}))\}^{\delta_{ij}}
\ \exp\left\{-W_i \ \lambda_0  \int_0^{y_{ij}}\ \Phi(b(s,{M}_i,\mathbf{x}_{ij})) \ ds\right\}\ , 
\label{eq:like}
\end{equation}
where $P[T_{ij}>t\mid W_i, M_i;\mathbf{x}_{ij}]=S(t\mid W_i,M_i;  \mathbf{x}_{ij}) 
=
\ \exp\left\{-W_i \ \lambda_0  \int_0^t\ \Phi(b(s,{M}_i,\mathbf{x}_{ij})) \ ds\right\}$,   
and the censoring indicator $\delta_{ij} =1$  
when observed $y_{ij}=T_{ij}$ and $\delta_{ij} =0$ when $y_{ij}<T_{ij}$.

\subsection{Spatial model for frailty:}
 To account for a spatial association among unknown frailties $\mathbf{W}=(W_1,\cdots, W_N)$ of $N=67$ counties (areal units) in (\ref{equ:model}), 
 the spatial distance between any pair $(i,k)$ of areal units (counties) is described by the $N \times N$ adjacency matrix $\mathbf{A}$ with $a_{ik}=1$ if $i\ne k$ are neighboring counties, otherwise $a_{ik}=0$. 
 We assume that the joint distribution of $\mathbf{R}=(\operatorname{log}(W_1),\cdots, \operatorname{log}(W_N))$ is a Conditional Auto-Regressive (CAR)  model \parencite{Besag74}, denoted by 
 \begin{equation}
\mathbf{R}\sim CAR(\mathbf{A};\sigma_1^2,\rho_1)\ \text{if}
\ \mathbf{R} \sim  MVN(\mathbf{0},\sigma_1^2(\mathbf{D}-\rho_1\mathbf{A})^{-1} )\ , 
\label{eq:CAR}
  \end{equation} 
  where $\mathbf{D}=\operatorname{diag}(a_{1+},a_{2+} \cdots a_{N+})$, and $ a_{i+}=\sum_{j = 1}^N a_{ij}$ is the total number of neighbors of county $i$. The unknown parameters $\rho_1\in (\frac{1}{\alpha_{(1)}}, \frac{1}{\alpha_{(N)}})$ and $\sigma_1^2>0$, respectively, characterize the spatial correlation and the variability among these $N$ cluster effects.
We need the restriction above on $\rho_1$ to ensure that the multivariate normal density of $\mathbf{R}$  in \eqref{eq:CAR} 
is proper, 
where $\alpha_{(1)}$ and $\alpha_{(N)}$ are the minimum and maximum eigenvalues of $\mathbf{A}$ and $(\frac{1}{\alpha_{(1)}}, \frac{1}{\alpha_{(N)}})=(-0.346,+0.163)$ for FL.
This spatial frailty model in (\ref{eq:CAR}) also ensures the identifiability condition $E[R_{i}]=0$,  and  the conditional distribution 
of each $R_i$ given $\mathbf{R}_{(-i)}=\{ R_j: j \neq i\}$ is $\mathrm{N} \left( \frac{\rho_1}{a_{i+}}\sum_{k = 1}^N a_{ik} R_k,\ {\sigma}_1^2/a_{i+} \right)$, where the conditional mean 
$E[R_i\mid \mathbf{R}_{(-i)}]$ is
proportional to the average $\sum_{k=1}^N a_{ik}R_k/a_{i+}$ of its neighbors' values, and the variance $\operatorname{Var}[R_i\mid \mathbf{R}_{(-i)}]$ is inversely proportional to the number of neighbors $a_{i+}$.

\subsection{Spatial model for unobservable covariates:} Instead of observing the true proportion of cluster-specific SMU ($M_i$) in FCR, another separate database, BRFSS, has the observed survey data $\mathcal{D}_0=(\mathbf{m}_0,\mathbf{n}_0)$ for $\mathbf{m}_0=(m_{01},\cdots, m_{0N})$ and $\mathbf{n}_0=(n_{01},\cdots, n_{0N})$, where $m_{0i}$ is the observed number of patients receiving regular screening mammography out of $n_{0i}$  patients surveyed in county $i$.
It is reasonable here to assume that the observed $m_{0i}$ given unknown $\mathbf{M}$ 
is independent $\operatorname{Binomial}(n_{0i},\ M_{i}) $, 
Also, 
where the spatial association model for $\mathbf{M}$ is similar to the spatial model (\ref{eq:CAR}) of $\mathbf{W}$ with the same adjacency matrix $\mathbf{A}$, but possibly with different spatial correlation $\rho_0\in (\alpha_{(1)}^{-1},\alpha_{(N)}^{-1})=(-0.346,0.163)$  and variability $\sigma_0^2>0$. 
Therefore, the likelihood contribution of $\mathcal{D}_0$ and the distribution $\mathbf{M}$ are given as
\begin{equation}
L_0(\mathbf{M}|\mathcal{D}_0)\propto \prod_{i=1}^N L_i^*(M_i)\propto \prod_{i=1}^N \{M_i^{m_{0i}}(1-M_i)^{n_{0i}-m_{0i}}\}, \ \text{and} \ \operatorname{logit}(\mathbf{M})\sim CAR(\mathbf{A};\sigma_0^2,\rho_0)\ , 
\label{eq:SMU}
\end{equation}
where $\operatorname{logit}(\mathbf{M})=  (\operatorname{logit}(M_{1}),\ldots, \operatorname{logit}(M_N))$.

To fully specify our hierarchical Bayesian model, we specify the joint prior distribution $p(\Omega)$ of our model parameters $\Omega=(b,\lambda_0,\eta_1,\eta_0)$ as $p(\Omega)\propto p(b \mid \Psi)\times 
p(\lambda_0)\times p(\eta_1)\times p(\eta_0)$,  where 
$p(\eta_1)$ is  the prior of the parameters $\eta_1=(\sigma_1^2,\rho_1)$ of the frailty model in  \eqref{eq:CAR}, $p(\eta_0)$ is  the prior of the parameters \(\eta_0=(\sigma_0^2,\rho_0)\)  of SMU in \eqref{eq:SMU}. 
As described earlier, we model the possibly time-varying effects, unknown forms of observed and unknown covariates $(\mathbf{x}_{ij}, M_i)$ and their interactions using the SBART prior $p(b \mid \Psi)$ on the completely nonparametric $b(t, M_i,\mathbf{x}_{ij})$, of \eqref{equ:model}, where $\Psi$ denotes its
usual regularized known hyperparameters of the priors of $\{\tau_K,\mathcal{M}_k\}_{k=1}^K$.  
We can show that the prior predictive survival function $E_{\Psi}[P\{T>t| W, M, \mathbf{x}\}]$ under SBART is a valid survival function, where $E_{\Psi}$ denotes the expectation with respect to the SBART prior.

Furthermore, for the independent priors $p(\eta_k)$ for $k=0,1$, we assume a common noninformative $\operatorname{Unif}(-0.346, 0.163)$ prior for both $\rho_k$, and a common independent Inverse-Gamma 
prior for both $\sigma_k^2$.
We use a Gamma distribution for prior \(p(\lambda_0)\). 
The hyperparameters for these priors are chosen to ensure that their choices have very moderate effects on the posterior inference, given the large size of the observed data.

\section{Posterior and computation}
Given the observed data  \(\mathcal{D}=\{\mathcal{D}_0,\mathcal{D}_1\}\), where \( \mathcal{D}_1= 
\{(y_{ij},\delta_{ij};\mathbf{x}_{ij}):j=1,\cdots,n_i;i=1,\cdots, N\}\) is the survival data from FCR and $\mathcal{D}_0=\{m_{0i},n_{0i}:i=1,\cdots, N\}$ is the mammography screening data from BRFSS, the joint posterior  of our model is
\begin{equation}
p(\Omega,\mathbf{W},\mathbf{M}\mid 
\mathcal{D}) \propto 
[\prod_{i=1}^N\{\prod_{j=1}^{n_i} L_{ij}\}]\times  L_0(\mathbf{M}|\mathcal{D}_0)\times g_1(\mathbf{W} \mid \eta_1)\times g_0(\mathbf{M} \mid \eta_0) \times p(\Omega) ,
\label{eq:posterior}
\end{equation}
where $L_{ij}$ is the likelihood contribution of the survival response of subject $(i,j)$ as given in \eqref{eq:like}, $g_1(\mathbf{W} \mid \eta_1)$ and $g_0(\mathbf{M} \mid \eta_0)$ 
are the CAR based joint densities of   $\mathbf{W}$ in \eqref{eq:CAR} and of 
$\mathbf{M}$ in \eqref{eq:SMU} respectively, and $L_0(\mathbf{M}|\mathcal{D}_0)$, a function of only unobservable $\mathbf{M}$, is the likelihood contribution from  BRFSS under the model of \eqref{eq:SMU}.
The first major challenge in sampling all the model parameters $\Omega=(b,\lambda_0,\eta_1,\eta_0)$ from \eqref{eq:posterior} using the MCMC tool is the presence of $\mathbf{M}$ in both $\prod_{(i,j)} L_{ij}$ and $L_0$. So, we use the following steps to implement the Bayesian computation of any Bayesian estimate $\int f(\Omega_1) p(\Omega_1|\mathcal{D})\ d\Omega_1$ of any parameter/quantity of interest in the model $f(\Omega_1)$ of the survival submodel in \eqref{equ:model}.

\noindent \textbf{Step~1}: Obtain samples of \(\mathbf{M}\) from the posterior $p_0(\mathbf{M},\eta_0\vert \mathcal{D}_0)\propto [L_0(\mathbf{M}|\mathcal{D}_0)]\times g_0(\mathcal{M}|\eta_0)\times p(\eta_0)$ 
and use these samples to compute  the posterior estimate \(\hat{\mathbf{M}}=E[\mathbf{M}| \mathcal{D}_0]\) of \(\mathbf{M}\). 

\noindent \textbf{Step~2}: Obtain MCMC samples of the rest of the parameters $\Omega_1=(b,\lambda_0,\eta_1)$ and $\mathbf{W}$  from the approximate posterior 
\begin{equation}
 p(\mathbf{W},\Omega_1 \mid \mathcal{D}_1;\hat{\mathbf{M}}) \propto 
[\prod_{i=1}^N\{\prod_{j=1}^{n_i} L_{ij}(W_i,\lambda_0,b;\hat{M}_i)\}] g_1(\mathbf{W} \mid \eta_1) p(b \mid \Psi) p(\lambda_0) p(\eta_1)\ . 
\label{eq:approxpos}
\end{equation}

\noindent \textbf{Step~3}: Use final samples of $(\mathbf{M},\eta_0, \Omega_1,\mathbf{W})$ from previous steps as importance samples from   \eqref{eq:posterior} with weights proportional to $\prod_{i=1}^N[\prod_{j=1}^{n_i}\{ L_{ij}(W_i,\lambda_0,b;{M}_i)/ L_{ij}(W_i,\lambda_0,b;\hat{M}_i)\}]$.

The primary challenge in implementing the MCMC tool to sample from \eqref{eq:approxpos} in Step~2 and evaluating the importance weight in Step~3  is that the likelihood contribution $L_{ij}$ in \eqref{eq:like} requires numerical evaluation of integrals like $\int_0^{y_{ij}} \Phi(b(s,\hat{{M}_i}, \mathbf{x})) \ ds$. In Step~2,  having this time-varying $b(s, \hat{M}_i,\mathbf{x}_{ij})$ inside the integral in \eqref{eq:like} prevents us from applying the usual Bayesian back-fitting algorithm (\cite{Chipman10}) to update the hyperparameters $\{\tau_k,\mathcal{M}_k\}$ of SBART given the rest of the parameters.
Following Basak et al. (2022), we address this using the Data Augmentation (DA) scheme with
additional latent variables $\mathbf{G}_{ij}=\{G_{ijk} \ \text{for}\ k = 1,\cdots,m_{ij}\}$ 
as the ``rejected" points 
generated from a Non-homogeneous Poisson Process (NHPP) with intensity
 \begin{math}
    \lambda_0 W_i  \{1 - \Phi(b(s, \hat{M}_i,\mathbf{x}_{ij}))\}
 \end{math}
 in the interval $(0, y_{ij})$, 
while viewing $y_{ij}$ as the first ``accepted" point of a thinned Poisson process
when $\delta_{ij}=1$. 
Given $(\Omega_1, \mathbf{W},\hat{\mathbf{M}})$ the conditional likelihood of the augmented $\mathbf{G}_{ij}$ simulated via this NHPP is given by 
 \begin{align}
     \begin{split}
        &\Pr[\text{
           ``Rejected" events at $\mathbf{G}_{ij}$, no other events in $(0, y_{ij})|\  \delta_{ij}$ ``accepted" event at $y_{ij}$ 
         }]
         \\
         & =
         \left[\prod_{k = 1}^{m_{ij}} 
         \{\lambda_0 W_i  (1 - \Phi(b(G_{ijk}, \hat{M}_i,\mathbf{x}_{ij})) \} 
       \right]   
       \exp\left[- \lambda_0 W_i\int_0^{y_{ij}}\{1-\Phi(b(s,{\hat{M}_i,\mathbf{x}_{ij}}))\}ds\right].
        \label{eq:augl}
     \end{split}
 \end{align}
Combining the augmented likelihood of (\ref{eq:augl}) with the observed likelihood $L_{ij}$ of \eqref{eq:like}, we get the ``complete data" likelihood contribution $L_{cij}(\lambda_0,b,W_i,\mathbf{G}_{ij}, {\hat{M}}_i)$
\begin{align}
  \propto\  \{ \lambda_0 
        W_i 
        \Phi(b(y_{ij}, \hat{M}_i,\mathbf{x}_{ij})\}^{\delta_{ij}}
      \times \prod_{k=1}^{m_{ij}}\left[
        \lambda_0
         W_i
        \{1-\Phi(b(G_{ijk}, \hat{M}_i,\mathbf{x}_{ij}))\}\right]
      \times 
      \exp(- \lambda_0 W_i  y_{ij}) 
      \label{equ:augl2}
    \end{align}
which is proportional to $\{\Phi(b(y_{ij}, \hat{M}_i,\mathbf{x}_{ij})\}^{\delta_{ij}}\prod_{k=1}^{m_{ij}}\{1-\Phi(b(G_{ijk}, \hat{M}_i,\mathbf{x}_{ij}))\}$ as a function of $b(\cdot)$. 
This allows us to use
the Bayesian back-fitting approach of Linero
and Yang (2018) for Binary response to sample $\{\tau_k,\mathcal{M}_k\}$ of $b(\cdot)$ from the joint posterior of (\ref{eq:approxpos}). This also ensures a closed-form Gamma 
conditional posterior for $\lambda_0$.

With the CAR model of $\mathbf{W}$ in (\ref{eq:CAR}), the resulting conditional posterior of $\mathbf{W}$ given $(\lambda_0,b,\mathbf{G}, \hat{\mathbf{M}})$ is not a standard density. Even via tools such as Metropolis-Hastings and Slice Sampling, it is challenging to sample from this density due to the large number of $N=67$ clusters, variable cluster sizes, and some really large clusters. 
To deal with such a complex conditional density of high dimension $\mathbf{W}$, we use Hamiltonian Monte Carlo (\cite{Duane87}), which uses the derivatives of the conditional density of $\mathbf{W}$ to generate an efficient candidate for the Metropolis step.

The MCMC samples of $(\mathbf{M},\Omega_0)$  and $(\mathbf{W},\Omega_1)$ a
obtained in Steps~1 and 2 are not the identically distributed samples from the true posterior in \eqref{eq:posterior} because the importance weight of $(\mathbf{M},\mathbf{W},\Omega_1)$ is proportional to $\prod_{i=1}^N[L_{ij}(\lambda_0,b,W_i, {M}_i)/L_{ij}(\lambda_0,b,W_i, {\hat{M}}_i)]$.
To avoid the challenge of evaluating the  $L_{ij}$ in \eqref{eq:like}, we instead use the MCMC samples  $\mathbf{M}$ and $(\mathbf{G},\mathbf{W},\Omega_1)$ from Steps~1-2 to obtain the sample $(\mathbf{G},\mathbf{W},\mathbf{M},\Omega_1)$ from the posterior \eqref{eq:posterior} in Step~3 with easy to compute weight proportional to $\prod_{i=1}^N[L_{cij}(\lambda_0,b,W_i,\mathbf{G}_{ij}, {M}_i)/L_{cij}(\lambda_0,b,W_i,\mathbf{G}_{ij}, {\hat{M}}_i)]$. 
Using sufficiently large number of samples of $(\mathbf{G},\mathbf{W},\mathbf{M},\Omega_1)$ and their corresponding normalized weights, we can obtain Monte Carlo approximations of any posterior prediction and posterior estimates of any function of interest $f(\mathbf{M},\mathbf{W}, \Omega_1)$ of the survival submodel in \eqref{equ:model}.  

\begin{algorithm}[t]
  \caption{MCMC algorithm to sample from posterior based on $(y_{ij},\delta_{ij})$.}
  \textbf{Input:} Observed censored survival times $(y_{ij},\delta_{ij})$; Posterior estimates $\hat{\mathbf{M}}$; Initial values of  $b$ (SBART), $\lambda_0$ (baseline hazard), $\mathbf{W}$ (frailties) and $\eta_1$ (frailty density parameter).

  \begin{algorithmic}[1]
  \For{$n^* = 1, \ldots, N^*$}
    \For{$i = 1,\ldots,N$}
      \For{$j = 1,\ldots,n_i$}
        \State $q_{ij} \sim Poisson(1; \lambda_0 y_{ij} W_i)$
        \State $\tilde{G}_{ij} \sim Uniform(q_{ij};  0,  y_{ij})$

        \State $\mathbf{U}_{ij} \sim Uniform(q_{ij};  0 , 1)$
        \State $\mathbf{G}_{ij} \gets \{\tilde{G}_{ij}: U_{ij} \le 1-\Phi(b(\tilde{G}_{ij}, \hat{M}_i,\mathbf{x}_{ij}))\}=\{G_{ij1}, \cdots, G_{ijm_{ij}}\}$

      \EndFor $\quad  j$
  \State Update $b$ using Bayesian back-fitting  (Linero and Yang, 2018)
  \State Update $\lambda_0$ by Metropolis-Hastings 
  \State Update $\eta_1$ by Hamiltonian Monte Carlo
  \State Update $\mathbf{W}$ by Hamiltonian Monte Carlo
    \EndFor $\quad i$
  \EndFor   $\quad n^*$
  \end{algorithmic}
  \label{alg:inference}
\end{algorithm}

\section{Simulation Study}
In this section, we compare the performance of our spatial SBART method with two existing competing methods: (1) Random Survival Forest \parencite{RandomSurvivalForest} without frailty;  and (2) the proportional hazards model with independent cluster-specific frailty (\cite{Hougaard95, Sargent}), called PH-frailty model now onwards. In both of these existing methods, we use the observed sample proportion $ \frac{m_{0i}}{n_{0i}}$ to augment the unobserved $M_i$. 
We compare these methods under 3 different Scenarios (given below) of the true hazard  
$\lambda_{ij}(t\vert W_i; M_i,\mathbf{x}_{ij})$, where each simulated dataset has   $n=2,000$ number of subjects, 2 continuous fully observed subject-level covariates  ($x_{1ij},x_{2ij}$) simulated from independent $\operatorname{Uniform}(0,1)$, and 1 cluster-level (considered unobservable) continuous covariate $M_i$ for $i=1,\cdots,10$ clusters of fixed cluster size of 200. 
The adjacency matrix $\mathbf{A}$ is based on 10 western counties of FL (to ensure some similarity to our motivating study). For each scenario, our observed data $\mathcal{D}_0$ for supplementary 
study is simulated as $m_{0i}\sim \operatorname{Binomial}(n_{0i},M_i)$,
where $n_{0i}$ is same as the observed $n_{0i}$ of the corresponding FL county. 
So, each simulated  data set $\mathcal{D}$ 
has observations $\{y_{ij}; x_{1ij}, x_{2ij}, n_{0i}, m_{0i}\}$ with only uncensored 
survival times $y_{ij}$ with $\delta_{ij}=1$.
For comparing 3 competing methods under each simulation scenario, the Monte Carlo approximations of the following criteria of the estimates are based on 20 replications. Each SBART based analysis used $2,500$ burn-in followed by $5,000$ MCMC samples.

    \textbf{Scenario~A} aims to compare the methods when the true hazard $\lambda_{ij}(t| W_i,M_i;x_{1ij},x_{2ij})= W_i \exp(t x_{1ij}+ 0.25x^2_{1ij}+0.5M_i)$ satisfies our non-linear non-proportional hazards model  in \eqref{equ:model}, except, one of the continuous covariates $x_{2ij}$ does not affect the hazard. In addition, both the random cluster effect $\mathbf{R}$ and the unobservable $\mathbf{M}$ follow the spatial CAR models of \eqref{eq:CAR} and \eqref{eq:SMU} with true $\sigma_0=\sigma_1=1$ and $\rho_0=\rho_1=0.5$. 
    Furthermore, to assess the robustness of the SBART method, the true hazard is chosen as exponential in time $t$, which is difficult to approximate using BART's piecewise constant hazard. 

\textbf{ Scenario~B} aims to compare the robustness of SBART 
with competing methods when the true common distribution of $\mathbf{W}$ and $\mathbf{M}$ is same as in Scenario~A, however, the frailty effect $W_i$ on the true hazard $\lambda_{ij}(t| W_i,M_i;x_{1ij},x_{2ij})= \exp(t x_{1ij} W_i+ 0.25x^2_{1ij}+0.5M_i)$ is not multiplicative as in \eqref{equ:model} due to its time-varying interaction $tx_{1ij}W_i$ with the covariate $x_{1ij}$. 

    \textbf{Scenario~C} aims to make another comparison of the robustness of these methods when the unobservable $M_i=\frac{W_i}{1+W_i}$ in the true hazard $\lambda_{ij}(t| W_i, M_i;x_{1ij},x_{2ij})= W_i \exp(t x_{1ij}+ 0.25x^2_{1ij}+0.5M_i)$ is a function of frailty $W_i$. 
    

The first criterion for comparing methods is the plot of the Monte Carlo approximation of $\mathcal{E}[\hat{S}(t|{\bf x}, \hat{M}_i)]$, 
called Average Estimated Survival (AES) curves, for different pre-specified values of $(x_1,x_2, M_i)$, where $\mathcal{E}$ is the sampling distribution of the estimators under the simulation model. For simulation scenario A, Figure~2 presents AES curves from 3 competing methods as well as the true survival curve for 4 different sets of values of $(x_1,x_2)$, 
but, for $M_i$ of one particular county. For all cases, the AES curve from the SBART method is clearly much closer to the true survival curve than other methods.
For other values of $M_i$ and also for simulation scenarios B and C, we found the AES curves from SBART methods to perform better than competing methods. 
The Figure S1 and S2 of Supplementary Materials present these true survival curve and AES curves from 3 competing methods respectively for simulation scenarios B and C (for two representative sets of $(\mathbf{x},M_i)$ for each scenario).

\begin{figure}

\begin{minipage}[t][0.3\textheight][t]{\textwidth}
  \includegraphics[width=1\textwidth]{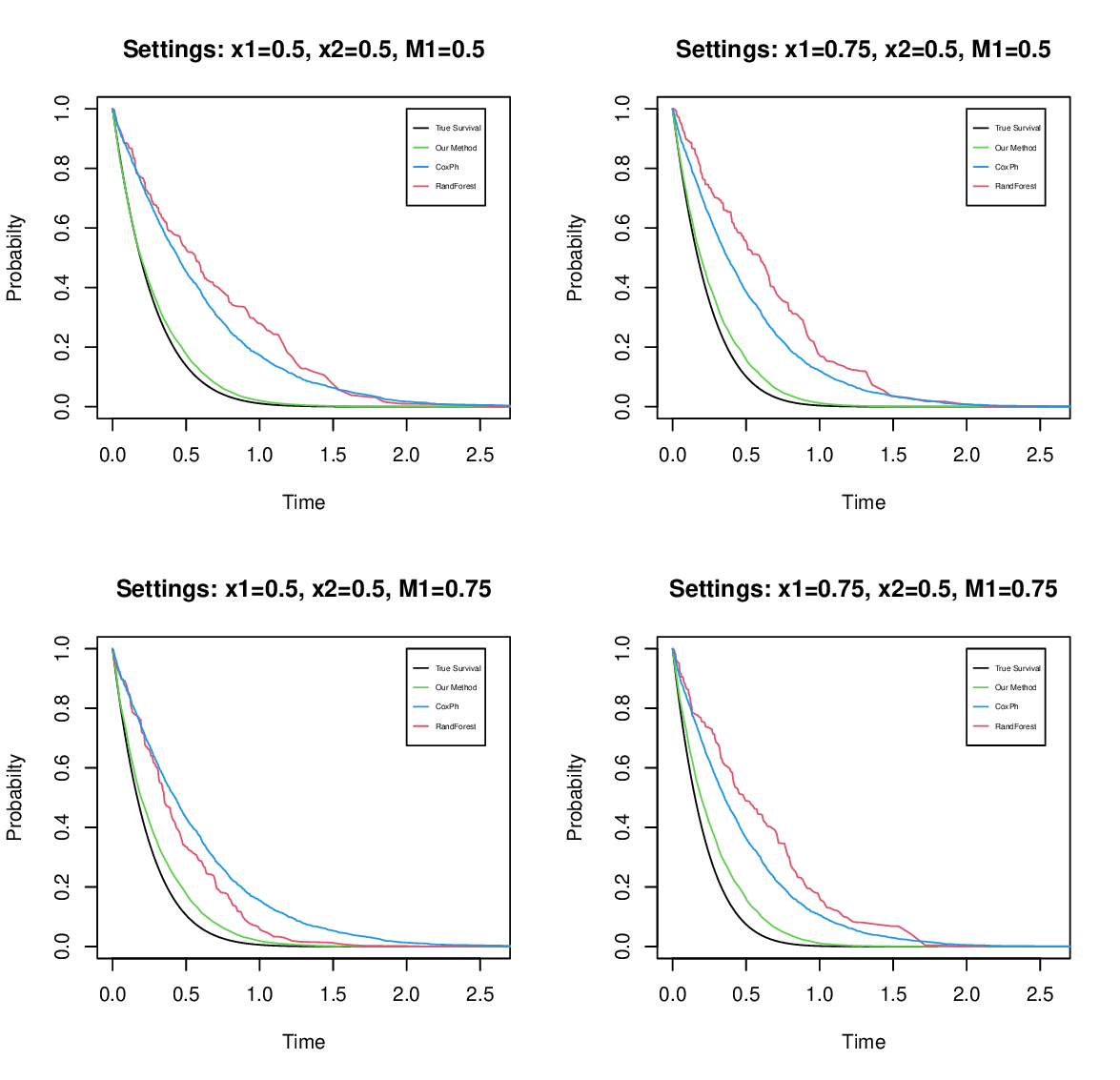}
\end{minipage}

\caption{\textbf{Simulation A:} Comparison of the Average Estimated Survival (AES) curves ($E[\hat{S}(t\mid\mathbf{x}_o,\hat{M}_i)]$) from 3 estimation methods and the true survival curve $S(t\mid\mathbf{x}_o,{M}_i)$ for $M_i$ of a particular cluster and 4 different values of $\mathbf{x}_o=(x_{1o},x_{2o})$. 
\textcolor{black}{ \textbf{--------}}: True survival curve; \textcolor{green}{ \textbf{--------}}: estimated by SBART;
\textcolor{red}{ \textbf{--------}}: estimated via Random Survival Forest; 
\textcolor{blue}{ \textbf{--------}}: estimated via PH frailty model.}
\end{figure}

The second performance metric for comparing estimation methods is 
$$
 \operatorname{AMSE} 
 =\frac{1}{N  t_{max}}\sum_{i=1}^N\int_{\mathcal{X}}
\int_0^{t_{max}} \mathcal{E}\{S(t|\mathbf{x}_{o},M_{i},W_{i})-\hat{S}(t|\mathbf{x}_{o},\hat{M}_{i},\hat{W}_{i})\}^2\ dt\ d\mathbf{x}_{o}  
 $$
(Average Mean Squared Error) of estimated survival functions $\hat{S}(t|\mathbf{x}_{o},\hat{M}_{i},\hat{W}_{i})$ from all $N=10$ clusters. Our Monte Carlo approximations of AMSE use $m=121$ pairs of $\mathbf{x}_{o}=(x_{1o},x_{2o})$ taken on a grid of the covariate space $\mathcal{X}$,  $G^*=150$ grid-points in the pre-specified time-interval of $0$ to $t_{max}=10$ years. 
SBART with spatial frailty has superior performance, that is, lower AMSE (0.25, 0.1, and 0.2 respectively for Scenarios~A, B, and C) than competing methods, with values (0.26 and 0.15 for Scenario~A and B) of PH frailty coming close second. Random Forest has the worst AMSE values, ranging between 0.43 and 0.49.
In conclusion, our method using SBART with spatial $\mathbf{W}$ and $\mathbf{M}$ outperforms competing methods even when the modeling assumptions of the multiplicative effect of $W_i$ (Scenario~B) and independence of $M_i$ and $W_i$ (Scenario~C) are not correct. 

\section{Application: Florida Breast Cancer (FCR) Study}
\label{s:sbart-psa_data_analysis}

For the FCR,
we consider 6 strata based on combinations of two races (African American (AA) and Non AA (WA)) and 3 stages of diagnosis (Distant/Regional/Local abbreviated as D/R/L). Important patient-level covariates include the continuous covariate age and categorical covariates Hormone Receptor (HR) status (negative/positive), tumor grade (1/2/3), Biopsy Delay or BD (long/short), and Treatment Delay or TD (long/short).
For all 3 stages of BC, the existing disparity in cancer survival for AA compared to WA at the national level is very well recognized \parencite{BCDisparity, ACS22}. Particularly for FCR registry from FL, this is very evident from the Kaplan-Meier plots given in Figure~S1 of the supplementary materials
and stratified log-rank test (p-value $<0.01$) to evaluate racial disparity in survival across all 3 cancer stages. 
For each of 6 strata, we use 
two competing survival analysis methods: our SBART method with spatial ($\mathbf{M},\mathbf{W}$) and the clustered survival SBART method of Basak et al.\ (2022). The second method use the same Bayes estimate $\hat{M}_i$ of $M_i$ from Step~1 as the plug-in for imputing $M_i$, however, does not use spatial association among $\mathbf{W}$. 
A comparison of these two methods based on Cox-Snell, as well as Martingale residuals (Figure~S4 in supplementary materials) shows that  our method 
performs better across all 6 strata. 

 We consider three non-demographic covariates (patient-level biopsy and treatment delays and county-level SMU) as ``intervenable" because the racial disparities in them can be addressed in the future by personal and policy-level interventions. The primary analysis goals are the following: (1) Identify the most critical covariates affecting the BC survival in all 6 strata (3 cancer stages $\times$ 2 races); (2) Determine the effects of these covariates within each stratum;  (3) For each county, assess the life-years saved (explained later) over 5 and 10 years in AA patients when the racial disparities in 3 intervenable covariates are mitigated.  

Table 1 presents the importance of 7 covariates based on the average number of times the particular covariate appears in any tree sampled from the posterior in (\ref{eq:posterior}). 
Let $c_{lm}$ be the number of splitting rules using the $l$-th predictor $x_l$ as the splitting variable 
out of the total $c_{.m}=\sum_{l=1}^p c_{lm}$  splitting rules in the $m$-th posterior sample. 
Using  $m=1 \cdots M^*$  posterior samples of BART,   
the importance of the predictor $x_l$ is measured by the Monte Carlo approximation $\nu_m=\frac{1}{M^*}\sum_{m=1}^{M^*} \frac{c_{lm}}{c_{.m}}$   (\cite{Luo14}).

\begin{table}[p]

\centering
\begin{minipage}{170mm}
  \caption{Posterior Importance Measures of 7 covariates within $2\times 3$ strata (2 races: AA for African-American, WA for rest; 3 cancer stages: L/R/D for Local/Regional/Distant). The top two important variables in each stratum are highlighted in boldface.}
  \smallskip
  \centering

\renewcommand{\tabcolsep}{6pt}

\begin{tabular}{cccccccc}
      \hline
           Strata: (Race, Stage) & &  (AA, L)&  (WA, L)&  (AA, R)& (WA, R)& (AA, D)&   (WA, D)\\ 
           \hline 
           Age & \vline &  0.105& \textbf{0.171}&  \textbf{0.163}& \textbf{0.165}& 0.135&     \textbf{0.173}\\
           HR Status &\vline &  0.140& 0.135& \textbf{0.178}& 0.122& 0.123&     0.131\\
           
           Tumor Grade  &\vline &  0.124& 0.146& 0.112& 0.146& 0.119&     0.146\\
           SMU & \vline &  \textbf{0.177}& \textbf{0.159}& 0.137& \textbf{0.152}& \textbf{0.165}&     0.144\\
           Treatment Delay & \vline &  \textbf{0.161}& 0.128& 0.126& 0.133& \textbf{0.168}&     \textbf{0.151}\\
           Biopsy Delay & \vline &  0.114& 0.145& 0.138& 0.145& 0.127&     0.131\\
           \hline
      \end{tabular}

\end{minipage}
\label{t:importance}
\end{table}

Unsurprisingly, Age is found to be important across most of strata possibly suggesting both hazard being a function of time  and time-varying effects of the covariates on hazard. 
For WA patients, biological covariates (such as Age, Tumor Grade) turn out to be most important. Whereas for AA patients, intervenable covariates of SMU and Treatment Delay (TD) are the most important covariates, especially for Local and Distant stage cancer patients. These results suggest that improving proportions of SMU and TD may lead to better survival rates particularly among AA patients.
The Boxplots of the within county sample proportions of TD, Biopsy Delay (BD),  and Bayes estimates $\hat{M}_i$ for SMU) in Figure~3 show the racial differences in these 3 intervenable covariates. We find that the Boxplots of the $\hat{M}_i$ of SMU and BD proportions have the biggest differences between two races.

\begin{figure}
    \centering
    \includegraphics[scale=.8]{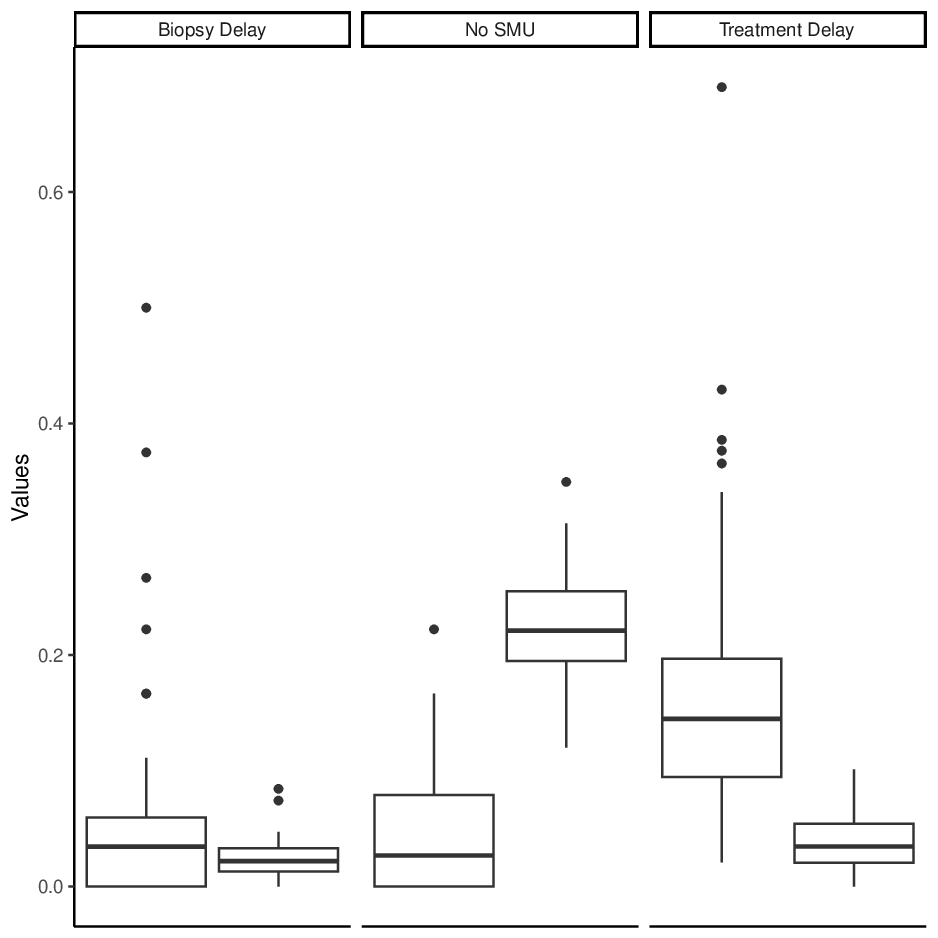}
    \caption{Boxplots of the proportions of patients having  Biopsy delay, estimated proportion of No SMU ($\hat{M}_i$), and proportions of Treatment delay (respectively from left to right in 3 panels). Left box: African American (AA), Right box: Not AA (WA).}
    \label{fig:my_label}
\end{figure}


A comprehensive analysis of the effects of these 3 intervenable covariates should look at their effects on disparities in all 67 counties and 3 stages separately, but, for sake of brevity, here we focus on only Broward county as an example because it is a big diverse county in FL and its SMU and BD proportions are very similar to the corresponding medians of 67 counties.  
For 6 strata, Figure~4 shows the estimated survival curves for representative patients in Broward county with 3 different stages of cancer. 
Here a 
 ``representative patient" for, say stage L and race AA, is a future AA and L stage patient from the county with covariate values $\mathbf{x}_{oi}$ same as the corresponding sample medians of all $L$ stage  AA patients within the same county. 
Especially for representative L and D stage patients in Broward, the estimated survival for AA (dashed line) is well below that of WA (solid line), with large difference in survival for stage D patients after year 1. According to our model, overall across all 3 stages, there is substantial racial disparity in survival between representative AA and 
WA patients in Broward.

\begin{figure}
    \centering
    \includegraphics[scale=.65]{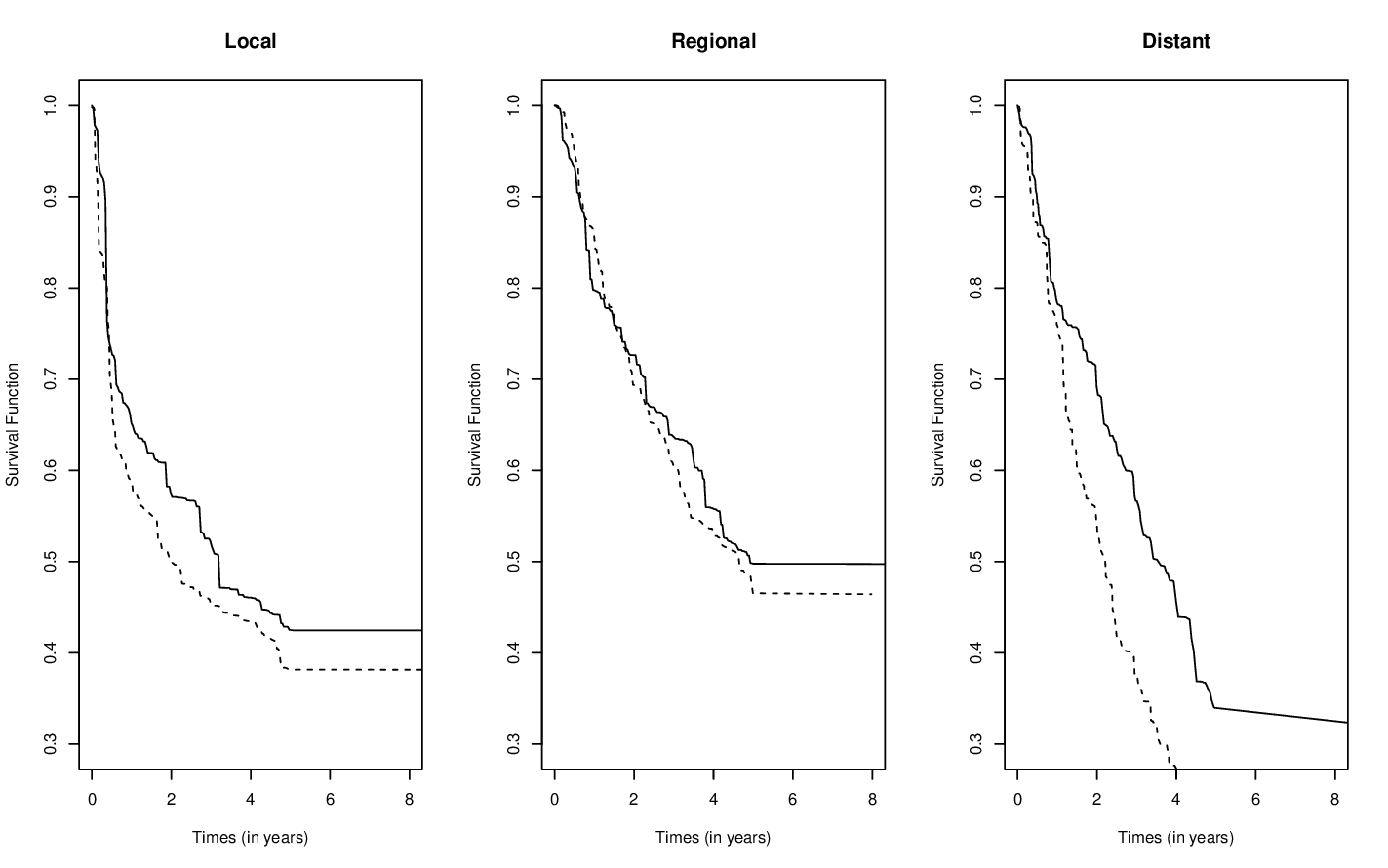}
    \caption{Estimated survival curve ($\hat{S}(t)$) versus time $t$ (in years) for AA (dashed line) and WA (solid line) patients with covariates values same as the observed median values within their respective racial groups in Broward county.}
    \label{fig:my_label}
\end{figure}

To understand the effect of mitigating disparity in 
any intervenable covariate of any particular AA patient$j$ at any particular diagnosis stage from county $i$, we need to compare the survival curves of two future AA patients with covariate values $\mathbf{x}_{oij}$ and $\mathbf{x}_{oij}^*$, where $\mathbf{x}_{oij}$ is the observed covariate of the AA patient $j$ from county $i$ and  $\mathbf{x}_{oij}^*$ is the same covariate value except a particular intervenbale covariate, say, BD, has been changed to no BD (best scenario to eliminate disparity). 
We use the metric of  expected Life Years Saved 
within $a$ years of monitoring (typically, $a=5$ and 10 years for us)  $\operatorname{LYS}(a)=\int_0^{a} [S(t|\mathbf{x}_{oi}^*, W_i)  -  S(t|\mathbf{x}_{oi}, W_i)]\ dt$ for 
the AA patient from county $i$  if the covariate values  $\mathbf{x}_{oi}$ of the patient are changed to $\mathbf{x}_{oi}^*$ while keeping the county-specific $W_i$ same.
For example, Figure~5 presents the $LYS(10)$ values versus corresponding county sample-sizes of R stage ``representative" AA patients  when only SMU proportions of these AA patients are changed to the maximum $\hat{M}_i$ of  AW patients (all the subject-specific covariates of "representative" AA patient are same as the observed median values within county) from 67 counties. 
The plot shows that a rural county such as Citrus has high dividend (in terms of LYS in 10 years) for mitigating disparity in SMU for an ``representative" AA patient, and a large urban county like Broward also has a moderately high (around 0.4 years) of expected LYS for mitigating disparity in SMU in the same ``representative" AA patient.  

\begin{figure}
    \centering
    \includegraphics[scale=.75]{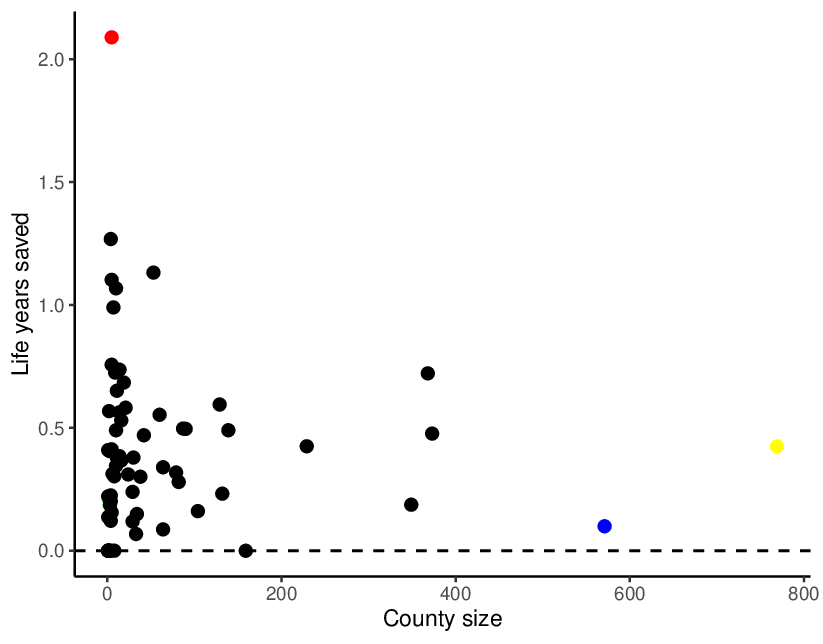}
    \caption{Plot of expected life years saved over 10 years for a representative AA patient with regional stage versus the respective observed sample sizes of 67 FL counties. \\ \tikzbulletgreen :  Baker County, \tikzbulletblue :  Escambia County, \tikzbulletyellow :  Broward County, \tikzbulletred :  Citrus County}
    \label{fig:my_label}
\end{figure} 

For R stage ``representative" AA patients (whose patient-specific covariates are same as corresponding observed medians for the county), Figure~5 illustrates the relationship between the $\operatorname{LYS}(10)$ if we mitigate the disparity in SMU in   67 counties versus 
the sample-sizes of the counties. This plot shows the variability among these counties and how some of the counties with small sample sizes can gain substantial in $\operatorname{LYS}(10)$.

\renewcommand{\tabcolsep}{20pt}
\begin{table}
\centering
\caption{Expected life-years saved over 10 years (LYS(10)) for a representative AA patient from Broward county if the racial disparities in three intervenable covariates are eliminated one at a time.}
\begin{tabular}{l|c|c|c} 
\toprule
Intervenable & Tumor & \multicolumn{2}{c}{\underline{Cancer Stage}} \\

Covariate & Grade & Local & Regional \\
\midrule
SMU & 1 &0.12 (0.11, 0.12) & 0.42 (0.41, 0.42) \\
& 2 & 0.16 (0.14, 0.16) & 0.43 (0.41, 0.43) \\
& 3 & 0.00 (0.00, 0.00) & 0.43 (0.42, 0.44) \\
\midrule
Biopsy Delay & 1 & 0.00 (0.00, 0.00) & 0.64 (0.64, 0.66) \\
& 2 & -0.02 (-0.02, 0.01) & 0.65 (0.65, 0.66) \\
& 3 & -0.11 (-0.11, -0.09) & 0.65 (0.65, 0.66) \\
\midrule
Treatment Delay & 1 &0.84 (0.84, 0.84) & 0.59 (0.59, 0.59) \\
& 2 & 0.83 (0.83, 0.83) & 0.60 (0.59, 0.60)\\
& 3 & 0.81 (0.81, 0.82) & 0.58 (0.56, 0.59) \\
\bottomrule
\end{tabular}
\end{table}

For each of 3 tumor grades and for L and R stage 
of an "representative" AA patients from Broward county, Table~2 presents the estimates (corresponding 95\% HPD interval in parenthesis) of $\operatorname{LYS}(10)$ when the disparity of the intervenable covariates (here, SMU, BD, or TD) are mitigated one at a time while keeping rest of them at their respective median values within Broward county.
For R stage AA patient, reducing disparity in BD seem have almost either no or small negative effects (HPD  are very very narrow covering/close to 0). 
These low and sometime negative effects can be 
artifacts of the association between BD and TD as well as very negligible effects of BD after adjusting for the estimated large effects of other important covariates. For example, removing SMU disparities results in up to 0.43 life-years saved for R stage AA patient. 
Whereas for R stage AA patients, mitigating disparity in BD results in around 0.65 life-years (widths of 0.02 year for HPD) saved. The most substantial and consistent (across both stages and all 3 grades) gains are seen by eliminating disparity in 
TD, with up to 0.84 life-years saved in R stage AA patients. This indicates that timely treatment is essential for improving survival and reducing disparities. In summary, the comprehensive analysis of $\operatorname{LYS}(10)$ from all 67 counties shows substantial variability in the relative importance of the intervenable covariates and the magnitudes of the these effects across different counties. This suggests the possible benefit of prioritizing these intervenable covariates differently for different counties.


 \section{Discussion}

This study highlights the utility of a nonparametric, spatially aware survival model to assess the impacts of closing disparities in modifiable covariates such as treatment delay, biopsy delay, and screening mammography utilization (SMU) across different counties. By leveraging this approach, we can estimate quantities like life-years saved by mitigating racial disparities in these covariates at both the patient and county levels. Our findings suggest that these estimates can provide valuable insights for targeted policy decisions and resource allocation. Specifically, closing gaps in SMU and treatment delays in counties with significant disparities could lead to substantial gains in survival for underserved populations. These estimates can inform customized budget allocations, targeted healthcare interventions, and help prioritize counties where policy changes may yield the most considerable improvements in public health outcomes.

There are some potential criticisms and weaknesses of our methods that will be addressed in future research. Our work does not incorporate the associations that exist among different covariates. Some of the covariates such as BD should not be beneficial, however, the estimated gain for mitigating disparity in BD turned out to negative for some strata (in spite of being very small in magnitudes). This can be addressed via allowing only a monotone adverse effects of BD (which is clinically very reasonable assumption) using mBART (an extension of BART). However, that requires additional computational challenges.  
One of the biological variables (the stage of diagnosis) is actually an intermediate outcome variable which is another issue not directly addressed in current analysis. We also avoid the issue of causal effects of the intervenable covariates even though it may be of practical interest here. 

\subsection*{Supporting Information}

Code for running the analysis and additional results are available in the supplementary web appendix.

\subsection*{Data Availability Statement}

The data that support the findings of this study are available from the Florida Cancer Registry. Restrictions apply to the availability of these data, which were used under license for this study. Data are available the authors with the permission of the Florida Cancer Registry.

\end{document}
a


\maketitle

\section*{1. Simulation Study Results}

For two sets of values of  $(\mathbf{x},M_1)$, Figures S1 and S2 present the true survival curve $S(t|\mathbf{x},M_1)$ of cluster $1$,
and Average Estimated Survival ($AES=\mathcal{E}[\hat{S}_1(t|\mathbf{x})$) curves of cluster $1$ from 3 competing methods: (1) our spatial SBART, (2) Random Survival Forest without frailty;  and (3) the proportional hazards model with independent cluster-specific frailty (called PH-frailty model). In both of these existing methods, we use the observed sample proportion $ \frac{m_{0i}}{n_{0i}}$ to augment the unobserved $M_i$. These figures highlight the superior performance of the SBART method compared to competing methods even when the model assumption about the effects of $M_i$ and $W_i$ on hazards of the true simulations models (Scenarios B and C) are not correct.

\begin{figure}

\begin{minipage}[t][0.3\textheight][t]{\textwidth}
  \includegraphics[width=1\textwidth]{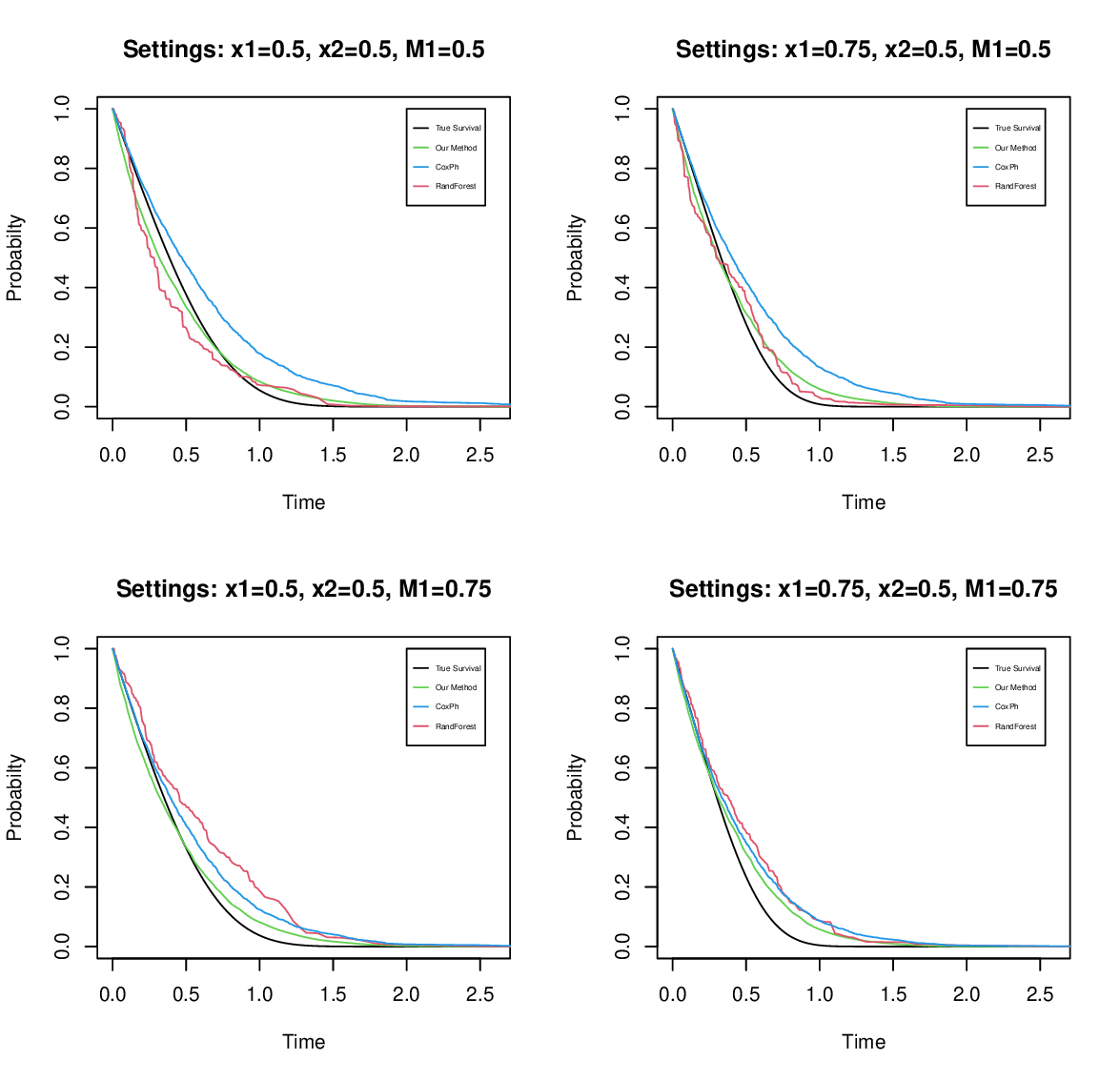}
\end{minipage}

\caption{Under \textbf{Simulation Model B}, comparisons of 3 estimation methods based on the Average Expected Survival curves $\mathcal{E}[\hat{S}_i(t\mid\mathbf{x})]$ of a cluster (cluster with true "unobserved" covariate $M_1$) under two different sets of true values of $(x_1,x_2,M_1)$. \textcolor{black}{ --------}: True survival curve\textcolor{green}{ --------}: Estimated by  spatial SBART;
\textcolor{red}{ --------}: estimated by Random Forest;
\textcolor{blue}{ --------}:  estimated by Hougaard's PH-frailty model (1995).
}
\end{figure}

\begin{figure}

\begin{minipage}[t][0.3\textheight][t]{\textwidth}
  \includegraphics[width=1\textwidth]{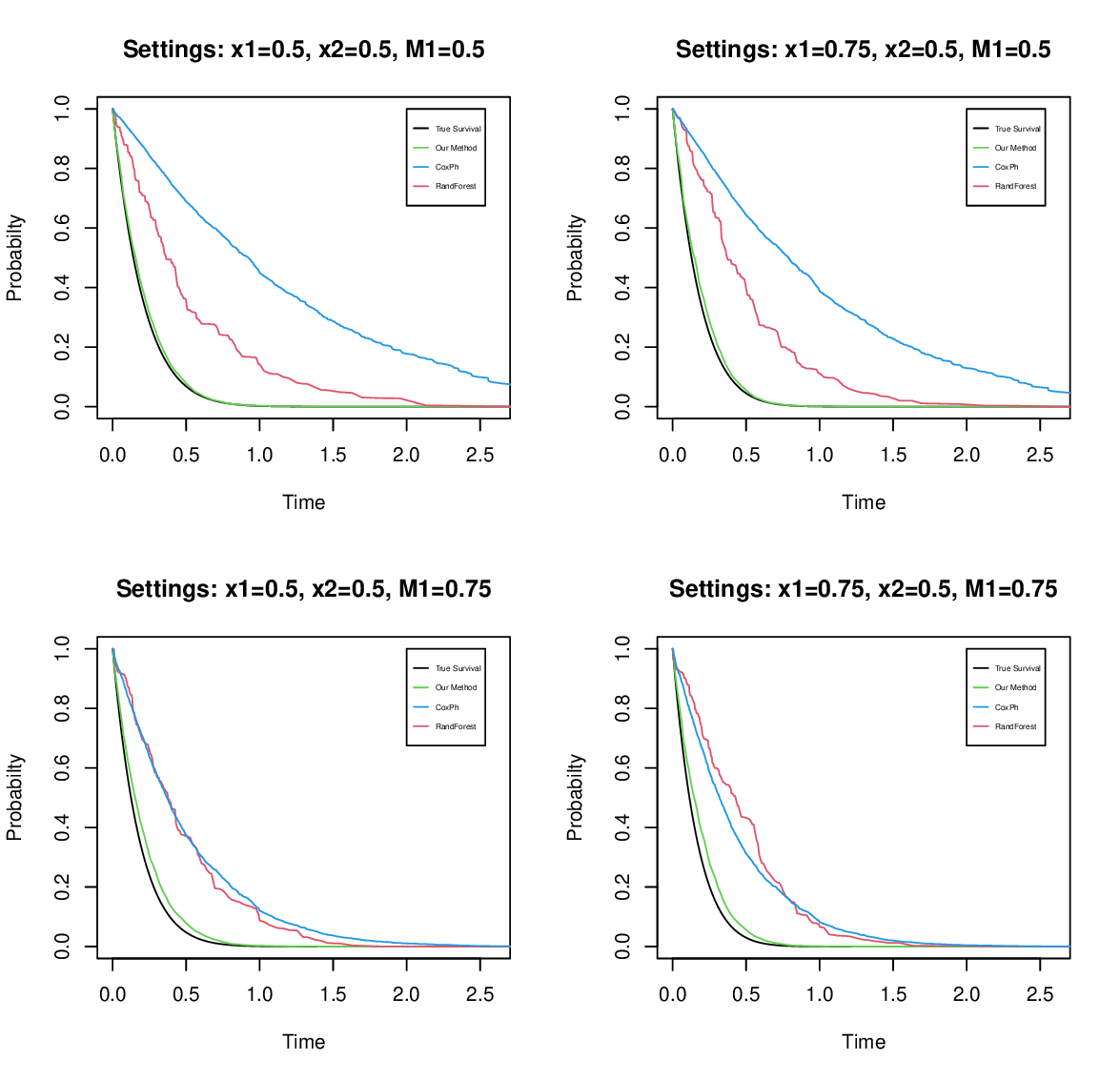}
\end{minipage}

\caption{Under \textbf{Simulation Model C},   comparisons of 3 estimation methods based on the Average Expected Survival curves $\mathcal{E}[\hat{S}_i(t\mid\mathbf{x})]$ of a cluster (cluster with true "unobserved" covariate $M_1$) under two different sets of true values of $(x_1,x_2,M_1)$. \textcolor{black}{ --------}: True survival curve\textcolor{green}{ --------}: Estimated by  spatial SBART;
\textcolor{red}{ --------}: estimated by Random Forest;
\textcolor{blue}{ --------}:  estimated by Hougaard's PH-frailty model (1995).
}
\end{figure}

\section*{2. Application: Florida Breast Cancer}

\subsection*{Kaplan-Meier Survival Plots}
Figure S3 shows Kaplan-Meier survival plots comparing African American (AA) and Non-African American (Non-AA) breast cancer patients, stratified by three stages of diagnosis: local, regional, and distant. These plots illustrate the disparities in survival rates between AA and Non-AA patients, with more pronounced differences observed in the distant stage. The results underscore the importance of addressing racial disparities in breast cancer treatment and survival outcomes.
\begin{figure}
    \centering
    \begin{minipage}[t][0.3\textheight][t]{\textwidth}
  \includegraphics[scale=.75]{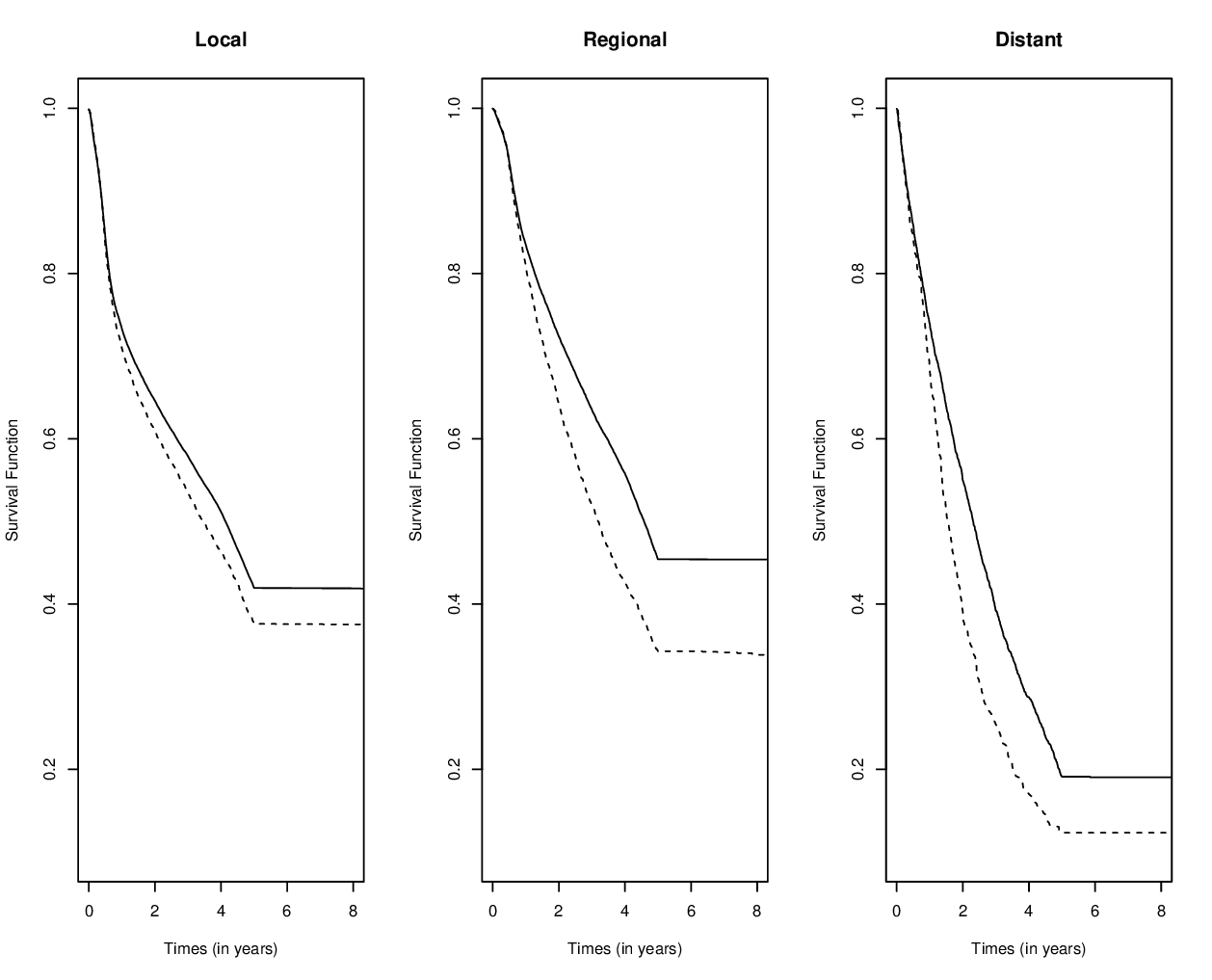}
\end{minipage}

    \caption{Kaplan-Meier survival plots for African American (AA: dashed line) and Non-African American (AW: solid line) patients, stratified by breast cancer stages (Local/Regional/Distant). The disparities in survival are most evident in the regional and distant stages, where AA patients consistently have lower survival probabilities. These results highlight the need for targeted interventions to address these racial disparities, within all 3 breast cancer stages.}
    \label{fig:my_label}
\end{figure}
\subsection*{Cox-Snell and Deviance Residuals}

Figures S4 and S5 present the Cox-Snell residuals and deviance residuals, respectively, comparing our spatial SBART method to the PH frailty model. These residual diagnostics help evaluate the goodness-of-fit of the models. The Cox-Snell residuals for both African American and Non-African American regional-stage patients (Figure S4) show better fit for our spatial SBART method compared to the PH frailty model. Similarly, Figure S5 shows deviance residuals for African American patients, demonstrating that our spatial SBART provides a better fit to the observed data especially for AA patients.

    \begin{figure}
    \centering
\begin{minipage}[t][0.3\textheight][t]{\textwidth}
  \includegraphics[scale=.65]{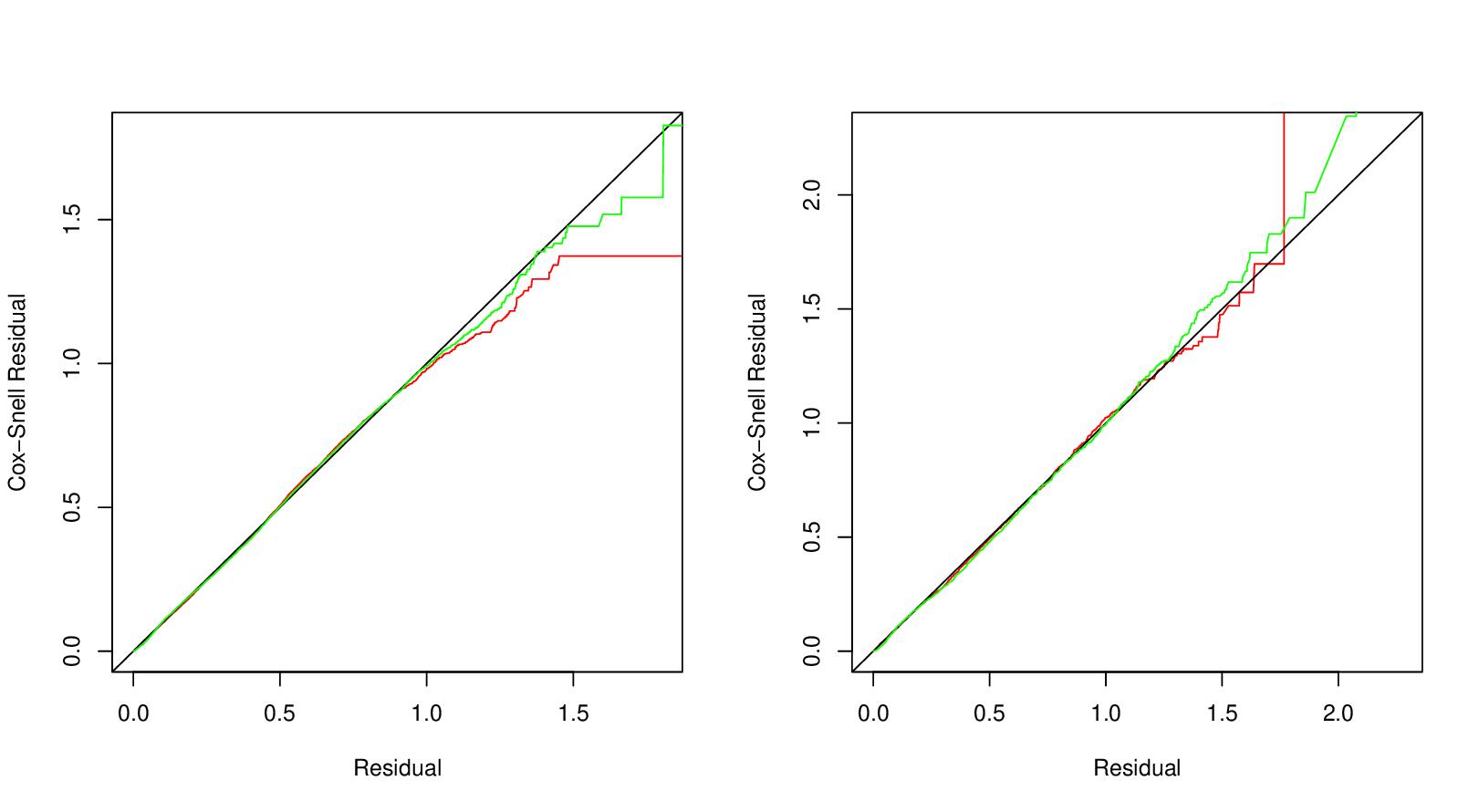}
\end{minipage}

    \caption{Comparison of our method (Green) to PH-frailty model (Red) of Hougaard (1995) by Cox-Snell Residual diagnostics for  for African American (AA) regional stage cancer patients (right panel) and non-AA (AW) regional stage cancer patients (left panel). }
    \label{fig:my_label}
\end{figure}

\begin{figure}[h!]
    \centering
    \begin{minipage}[t][0.3\textheight][t]{\textwidth}
 \includegraphics[scale=.7]{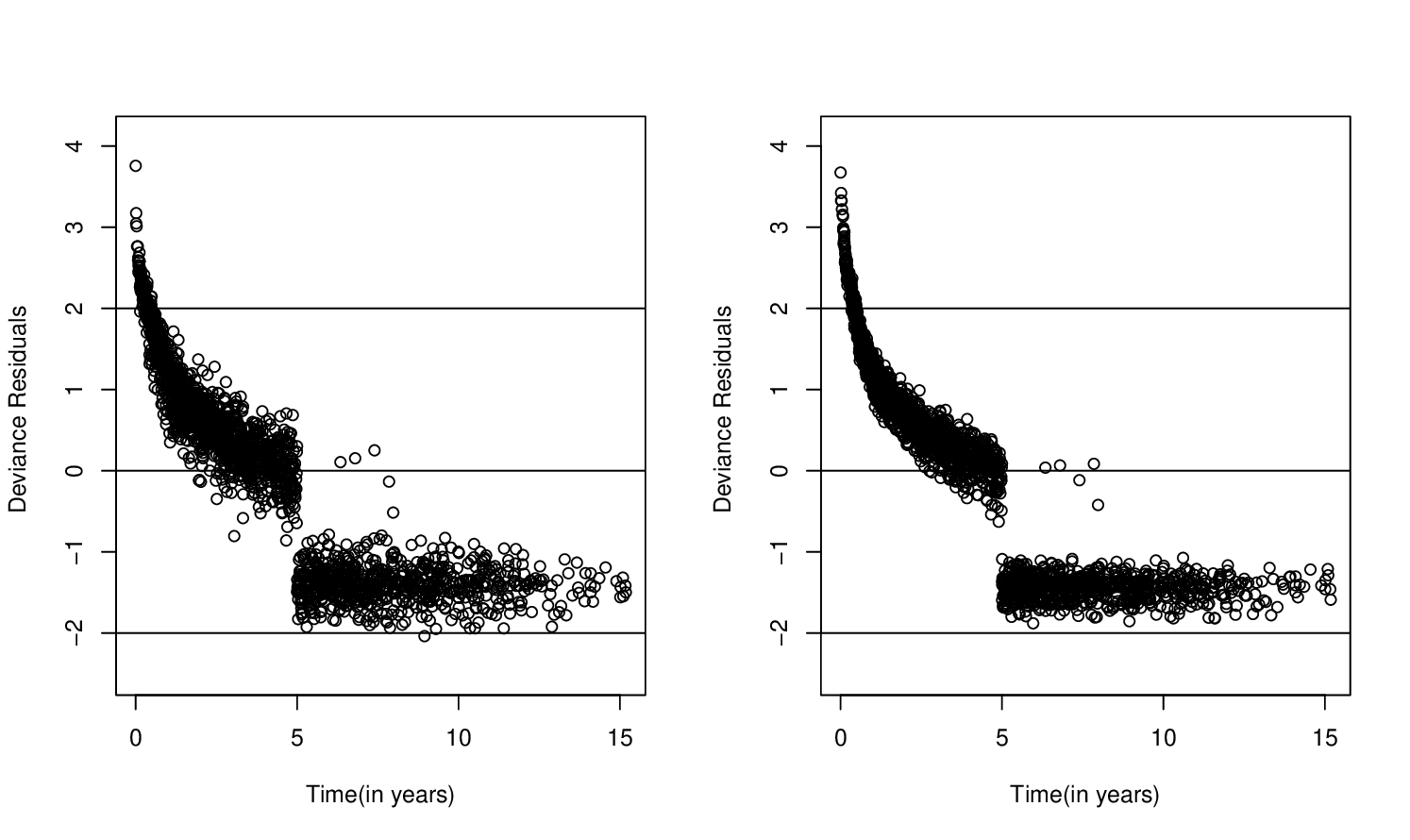}
  
\end{minipage}
   
    \caption{Comparison of our method (left) to Frailty model of Hougaard (right) by Deviance Residual diagnostics of African American  Cancer regional stage cancer patients}
    \label{fig:my_label}
\end{figure}

\begin{figure}
    \centering
      \begin{minipage}[t][0.3\textheight][t]{\textwidth}
  \includegraphics[scale=.7]{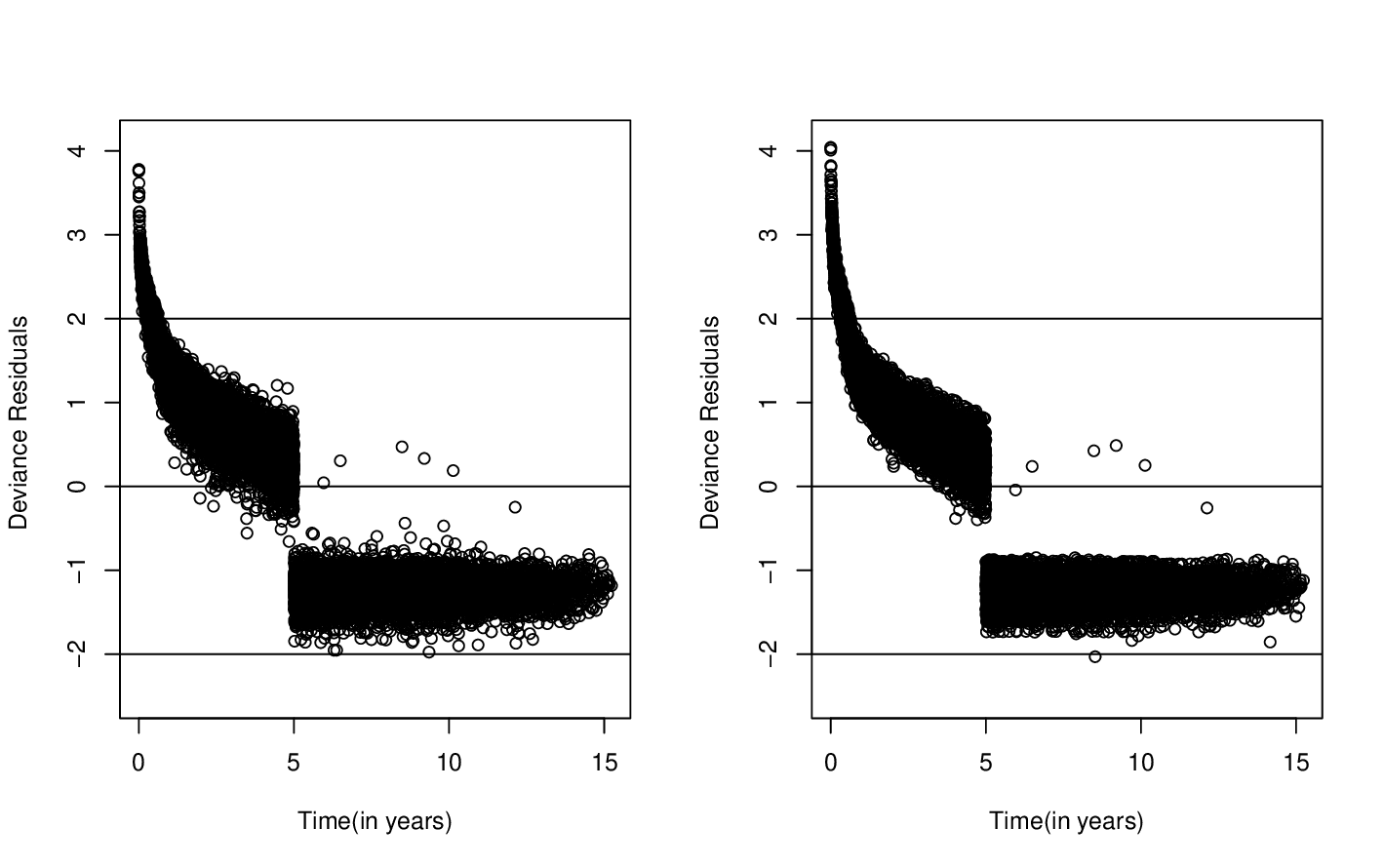}
  
\end{minipage}
   
    \caption{Comparison of our method (left) to Frailty model of Hougaard (right) by Deviance Residual diagnostics of American White regional stage cancer patients }
    \label{fig:my_label}
\end{figure}

\subsection*{Life-Years Saved Analysis}
Table S1 presents the estimated life-years saved over 5 years for African American patients in Broward County if racial disparities in Biopsy Delay (BD) and Screening Mammography Utilization (SMU) are eliminated. The analysis shows that improving SMU and avoiding BD can significantly improve survival outcomes, particularly for Regional-stage AA cancer patients. The table quantifies the potential benefits of these interventions, showing how targeted policies may improve survival of AA cancer patients in various FL counties by addressing disparities in cancer detection and care among AA patients.
\begin{table}[ht]
\centering
  \caption{ Expected life-years saved over 5 years for a representative AA patient from Broward county if the disparities in the BD and the SMU proportions are eliminated compared to AW cancer patients in other counties. }
        \begin{minipage}[t][0.3\textheight][t]{\textwidth}
 \begin{tabular}{rrrrr}
  \hline

& & Stage Local & Stage Regional & Stage Distant \\ 
  \hline\\

       &    Tumor Grade 1 &  0.304&  0.664&  0.021\\
   Biopsy  &      Tumor Grade 2&  -0.012&  -0.153&  0.144\\
        &   Tumor Grade 3&  0.028&  0.213&  -0.515\\
   \hline \\
 & Tumor Grade 1 &  0.458&  -0.076&  0.587\\
 SMU &         Tumor Grade 2&  0.672&  -0.279&  0.582\\
   &        Tumor Grade 3&  0.559&  0.200&  0.871\\
   \hline

\end{tabular}
\end{minipage}

\end{table}

\begin{figure}
    \centering
          \begin{minipage}[t][0.3\textheight][t]{\textwidth}
   \includegraphics[scale=.85]{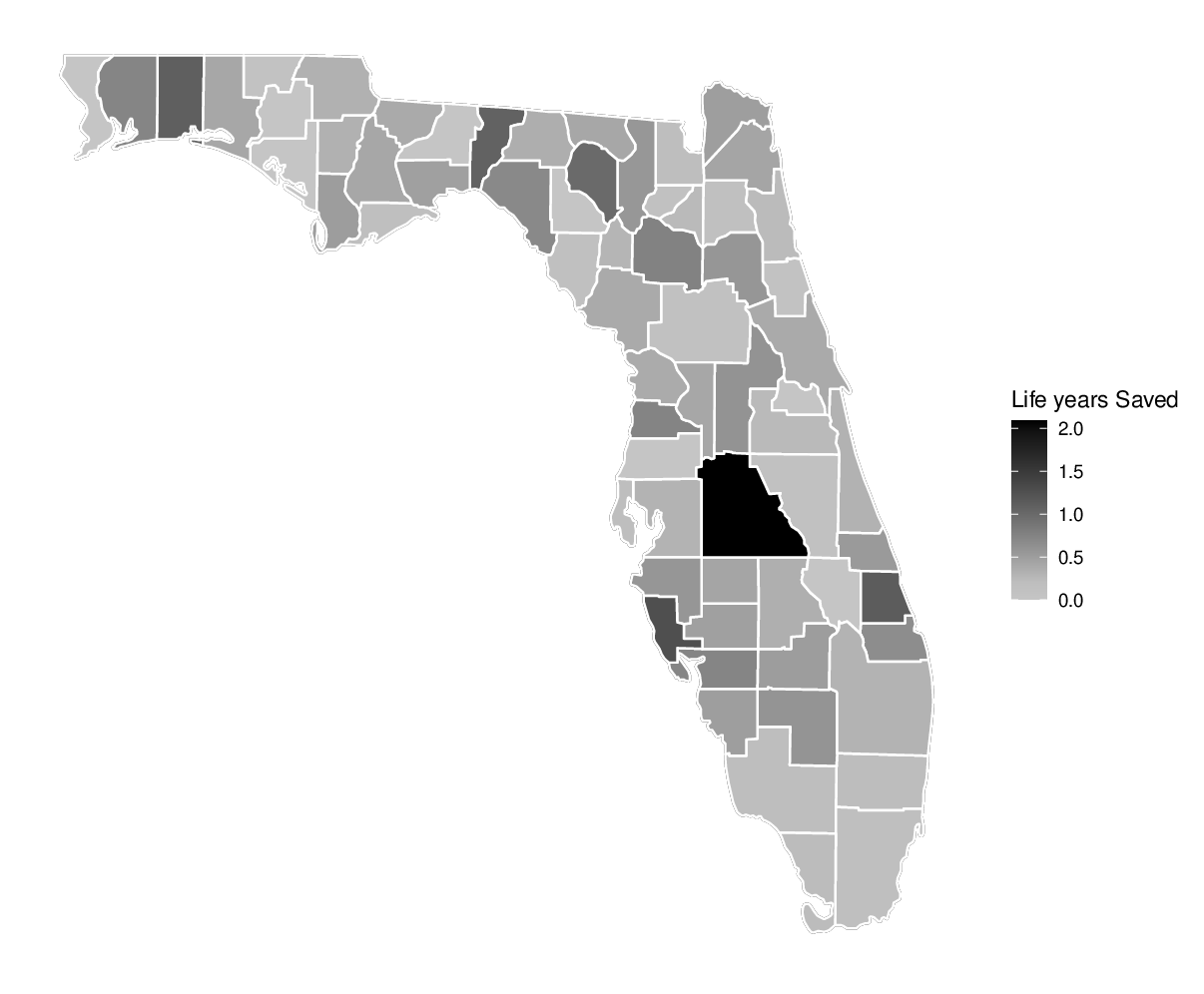}
\end{minipage}

    \caption{Expected life-years saved in 10 years for  "representative" African American patients of each FL county if the racial disparity in only SMU proportion is mitigated for each county.}
    \label{fig:my_label}
\end{figure}